# Magnetic ordering in metal-free radical thin films

Tobias Junghoefer,[1,†] Arrigo Calzolari,[2,†] Ivan Baev,[3,†] Mathias Glaser,[1] Francesca Ciccullo,[1] Erika Giangrisostomi,[4] Ruslan Ovsyannikov,[4,] Fridtjof Kielgast,[3] Matz Nissen,[3] Julius Schwarz,[3] Nolan M. Gallagher,[5] Andrzej Rajca,[5] Michael Martins,[3] Maria Benedetta Casu[1,*]

[1]Institute of Physical and Theoretical Chemistry, University of Tübingen, 72076 Tübingen, Germany

[2]CNR-NANO Istituto Nanoscienze, Centro S3, 41125 Modena, Italy

[3]Department of Physics, University of Hamburg, 22761 Hamburg, Germany

[4]Helmholtz-Zentrum Berlin, 12489 Berlin, Germany

[5]Department of Chemistry, University of Nebraska, Lincoln, United States

*benedetta.casu@uni-tuebingen.de

[†]These authors contributed equally to this work

**Magnetism in organic materials is very intriguing: the realization of long-range magnetic order in completely metal-free systems means that magnetic moments are coupled to useful properties of organic materials, such as optical transparency, low-cost fabrication, and flexible chemical design. Magnetic ordering in purely organic thin films is unknown so far since most of the investigations show this effect due to the proximity of light atoms to heavy metals, impurities, or vacancies. Here, we unravel X-ray magnetic circular dichroism at the carbon and nitrogen K-edges in purely organic radical thin films. Our results show a different behaviour depending on preparation than the magnetic signature of the single crystals. Atomistic simulations indicate that the reason for this is the molecular arrangement in the films compared to that in the single crystal. Our work opens a new avenue towards understanding magnetic properties in purely organic materials. The tuning of the film magnetic properties by the molecular arrangement is an exciting perspective towards revealing new properties and applications.**

Magnetic ordering in light elements, such as nitrogen and carbon, has been studied in magnetic-edged graphene nanoribbons[1] or bilayers[2], and polymers[3] while in hybrid organic thin films most of the investigations show this effect as due to the proximity of light atoms to heavy metals, impurities, or vacancies[4]. Purely organic radicals are molecules that carry one unpaired electron giving rise to a permanent magnetic moment, in the complete absence of metal ions.[5-7] Inspired by their tremendous potential, here we investigate thin films of an exceptionally chemically stable Blatter radical derivative[8] by using X-ray magnetic circular dichroism (XMCD)[9-12]. Here we observe XMCD at the nitrogen and carbon K-edges in the thin films: our results show a magnetic behaviour different than in the single crystals depending on the preparation conditions.

Purely organic radicals are exceptionally promising in various fields. Consequently, the needs arose to go beyond the limit of the first attempts to grow films by wet preparations that is prone to the presence of contaminants. The controlled radical thin film growth using evaporation was believed to be not practicable because of their reactivity that would cause their degradation during evaporation[13]. Once proved that several families of radicals are stable enough to be evaporable,[14] scientific curiosity and potential applications are driving the research on their thin film,[15-18] identifying their potential use in applications ranging from quantum computing to organic spintronics[16,17,19]. While several radicals are synthesized with high enough chemical stability to stand evaporation, their film forming properties are often very poor because of their high vapour pressure at room temperature and the low molecular weight that make them highly volatile and characterized by very low sticking coefficients[8,20]. Willing to investigate the magnetic behaviour of radical thin films, we faced several challenges. To achieve reproducible films and, thus, reproducible results, controlled conditions during growth are necessary to avoid any artefacts (contaminants, degradation due to film aging and air-exposure). The investigations require high signal sensitivity: a single magnetic moment is associated to the large molecular volume. The high sensitivity is furthermore necessary because we investigate very thin films (the nominal thickness varies in the 4-9 nm range). An important issue is also achieving very low temperature, simultaneously to the presence of high magnetic fields to reach magnetic saturation. In fact, purely organic radicals are predicted to have very low Curie temperatures[5-7,21]. We selected a chemically super stable Blatter radical[22,23] derivative, (Blatter-pyr, Figure 1a)[8] explicitly designed to have very good film forming properties and film stability.[8] XMCD was the experimental technique of our choice. It is an element-specific technique based on the absorption of polarized X-rays due to electron transitions from core levels to unoccupied states, a well-established method to access spin and orbital angular momenta in transition metals, identifying their magnetic behaviour [9-12]. Despite the XMCD well-known capabilities, the measurements of metal-free organic films required exceptional facilities to overcome all challenges. The experiments were performed using a newly developed ultra-low temperature setup equipped with a superconducting vector magnet allowing magnetic fields up to ±7 T in the horizontal direction and ±0.5 T in the vertical direction, and a cryogen free $^3$He-$^4$He dilution refrigerator, allowing experiments reaching the lowest temperature of 100 mK.[24] This unique portable setup features an ultra-high vacuum (UHV) preparation chamber that allowed growing the radical films by organic molecular beam deposition under controlled conditions, and transferring them to the measuring chamber without breaking the UHV. This led to the investigation of thin films free from contaminants because catalysts, solvents and air exposure may affect the intrinsic film properties.

**Results and Discussion**

We applied a magnetic field at very low temperature, to align the magnetic moments in the direction of the field that was collinear to the incoming circularly polarized photon beam. The absorption spectra (XAS) once with left and once with right circularly polarized light were measured (Figure 1b). XMCD is defined as the difference between the two absorption signals. We focus on the N K-edge absorption spectra measured at 1.1 K with a magnetic field of +7 T. The films in Figure 2 (nominal thickness 9 nm) were prepared by using controlled evaporation keeping the substrate at 290 K. We observe a clear dichroic signal (Figure 2). The signal reverses sign upon reversal of the magnetic field (Supplementary Information, Figure S1) confirming that the dichroism is real (magnetization reversal is equivalent to switching the helicity of the polarized light). We also measured the absorption at the C K-edge: a XMCD signal is visible also in this case. This behaviour is surprising because the Blatter-pyr single crystal is antiferromagnetic[25], in which case no XMCD should be observable. Discrepancies in the intensity of the XMCD for the two signs of the magnetic field suggests the presence of uncompensated spins that do not follow the external magnetic field[26-28]. Also, the curves are normalised considering the total absorption of the beamline (see the experimental section for details), as it is usually done for XMCD[10,29,30]. This method cannot compensate for possible substrate contaminants contributing to absorption.

To understand the origin of this unexpected behaviour we deposited and investigated a second set of Blatter-pyr thin films with different thicknesses (4 and 9 nm, the latter as in the measurements in Figure 2), at 0.15, 10 and 15 K in order to span a larger temperature range. We prepared the samples keeping the substrate temperature slightly higher (300 K). This increased the kinetic energy of the Blatter-pyr on the substrate during film growth.[8] We measured the X-ray absorption spectra of the films as previously done and we could not detect any XMCD (Figure 3). Also, we did not detect any dichroism spanning a large temperature range from 0.15 to 15 K. This result is concomitant with the antiferromagnetic behaviour of the Blatter-pyr crystalline bulk. Therefore, we found that, depending on the growth conditions, a few samples resemble the crystalline character, others a spin-uncompensated magnetic behaviour. We note that the different preparation conditions cause not only a different XMCD but also the presence of a new feature (Figure 3, and S3 in the Supplementary Information) when the samples are grown on the substrate at 300 K indicating differences in the electronic structure, expected when a different molecular arrangement occurs[31,32].

Thin films can have considerably different structural, transport, electronic and magnetic properties with respect to their bulk counterpart.[33,34] Polymorphism is particularly enhanced in organic materials because of the degree of freedom of the molecules in a film. Molecules have 3N degrees of freedom with N being the number of atoms in a molecule. This has pronounced consequences for all properties of the system. For example, molecular packing has been invoked to explain the different magnetic behaviour ranging from antiferro- to ferromagnetic[21,35] in the different crystallographic phases of the crystals of the first purely organic ferromagnet ever reported, the *p*-nitrophenyl nitronyl nitroxide radical[7]. In our experiments we observe that different preparation conditions play a role: decreasing the kinetic energy of the molecules on the substrates influences their adjustment in the films[33,36,37] with clear consequences on the magnetic interactions, since the antiferromagnetic character of the Blatter-pyr crystals depends on a delicate adjustment of the slippage angle[38,39].

Energetically, the antiferro- (AF) and the ferromagnetic (FM) interactions in the bulk are very close: we performed first-principles density functional theory calculations for the Blatter-pyr derivative that indicate that the antiferromagnetic configuration is more stable by only ~2 meV/cell. The electronic and magnetic properties of the crystalline bulk along with those of the single Blatter-pyr molecule are summarized in the Supplementary Information (Figure S4 and S5). The minimization of electronic problem for a non-magnetic state (para or diamagnetic) does not converge to a minimum, that means that the zero magnetization is not a (either total or local) minimum for the total energy, indicating that the non-magnetic phase is not the ground state for the system.

To shed light on the correlation between structural changes, induced by different preparations, and the intermolecular interactions, we have modelled the thin film phase. Starting from the 3D crystalline structure[8] (Figure 4a), we considered a 9 nm-thick film grown along the *c*-axis, as shown in Figure 4b. The film structure stems from the stacking of the four radicals (labelled 1-4 in the following analysis), which form the crystal. As for the crystalline case, the AF and the FM interactions are almost energetically degenerate. Hereafter, we assume the AF as the reference.

The total magnetic moment of the antiferromagnetic case is the sum of the single molecule contributions with antiparallel spin orientation. The antiparallel magnetic arrangement can be distinguished by the non-magnetic spin unpolarized case, through the evaluation of the absolute magnetization $\mu_A = \int |\rho_{up} - \rho_{dw}| d^3r$ . Non-magnetic systems have $\mu_T = \mu_A = 0$, while antiparallel spin systems have $\mu_T = 0$ and $\mu_A \neq 0$. In the present case $\mu_T$=0.0 $\mu_B$ and $\mu_A$=15.2 Bohr magneton ($\mu_B$)/cell. The spin alternation follows the structural bi-layer arrangement (↑↑ - ↓↓): spin-up for molecules 2 and 3, spin-down for 1 and 4, as shown by the charge spin-density plot (i.e., $\rho_{up}$-$\rho_{dw}$) in Figure 4a, where green and blue lobes represent spin-up and spin-down orientation, respectively. This reflects on the density of states (DOS, Figure 4b) that closely derives from the coherent superposition of the molecule spectra, where each radical exhibit an unpaired single occupied (SOMO, $S_O$) and the single unoccupied (SUMO, $S_U$) molecular orbitals[40]. For molecules 2-3, $S_O$ and $S_U$ have a spin-up and spin-down polarization respectively, while the opposite order attains for molecules 1-4. The electronic and magnetic properties of the film do not depend on the molecular orientation of the external layers (see Supplementary Information, Figure S6a). Similar results can be obtained also from films grown along the *a*- and *b*-axis respectively, see, e.g., Figures S6 b and c.

A closer inspection of the electronic properties indicates that, albeit weak, there is a long-range intermolecular coupling in the film. This is evident, e.g., from the band structure (Supplementary Information, Figure S7a): (i) the molecular-derived bands are not fully degenerate, but rather split in a 2-fold manifolds; (ii) the $S_O$ and $S_U$ bands have k-dispersion of ~100 meV across the Brillouin zone. $S_O$ and $S_U$ states, which are responsible for the magnetic character of the radical, are π-orbitals with a net contribution from both carbon and nitrogen atoms. Even though the magnetic origin of the radical is formally due to the presence of an undercoordinated N atom, the spin-density distribution (i.e., $\rho_{up}$-$\rho_{dw}$) is spatially delocalized all over the molecule and not centred on the nitrogen site[19]. Near edge X-ray absorption fine structure (NEXAFS) spectra mirror this delocalisation (see Supplementary Information Figure S8 and S9 (b and d)). This also explains the reason why we observed a XMCD signal both at the N and C K-edges in the films in Figure 2.

To unravel the interplay between the molecular packing and the magnetic character we consider a simple model constituted of a 3 nm-thick film, resulting from the stacking of the four molecules 1-4 (Figure 5a). Starting from the crystalline arrangement (θ=0°) we modified the molecular packing by changing the herringbone angle by 5°, 10°, 15° and 20°, with respect to the *c*-axis. The higher is the rotation angle the flatter is the molecular stacking. After full atomic relaxation we recognize very different configurations, as a function of the initial angles. For θ≤10° the systems maintain the structural and electronic characteristics of the 9 nm-thick films and the crystalline bulk, both AF and FM interactions can be obtained at the same energy. For $\theta \geq 15°$, the molecule-molecule interaction becomes predominant and the structure undergoes a strong spatial redistribution, as shown in Figure 5b for the case of $\theta=20°$. The external molecules 1 and 4 slightly displace from the centre of the film and rotate back restoring the crystalline ($\theta=0°$) spatial distribution. The internal molecules 2-3 rotate in the opposite direction (i.e. increasing $\theta$) forming a very ordered and close-packed bi-layer of parallel molecules. The resulting configuration has a total magnetic moment $\mu_T=\mu_A=2.0$ $\mu_B$/cell. This value is not the result of a transition from AF-to-FM interactions (the FM configuration would have $\mu_T=\mu_A=4.0$ $\mu_B$/cell), but rather of a non-complete spin compensation. Indeed, the charge spin density plot (panel 5b) shows the cancellation of the magnetic moment of the central molecules (2-3). The comparison of the magnetic moment per atoms in the limiting cases for $\theta=0°$ and $\theta=20°$ (Figure S5) indicates an identical behaviour for external molecules 1-4, and the complete quenching of the magnetic moment, mostly on N atoms, of the two central molecules, which is a fingerprint of a local charge redistribution. As mentioned above, SOMO and SUMO are aromatic $\pi$ states, delocalized over the entire molecules including nitrogen atoms. The close packed configuration of molecules 2 and 3, caused by the rotation, favours the $\pi-\pi$ coupling, which results in the constructive hybridization of the molecular states, to form two new mixed delocalized orbitals across the film (Figure 5c). The intermolecular interaction changes the DOS plot of the system (panel d), where the external molecules (1-4) maintains their original spin-polarized character as in the thick film, while the internal layer gives rise to a spin-unpolarized peak that crosses the Fermi level (M-state in panel 5d and in the band structure of Figure S7b).

The modification of the magnetic properties caused by a different molecular arrangement is the key to understand the differences observed in the experimental results. We have also considered the possible effects of structural changes (e.g., flattening of the molecules in the proximity of a substrate), which may quench the magnetic moment of the first layers. We changed the molecular arrangement of the bottom part of the 9 nm-thick film, by rotating only the lowest 4 molecules by 20°, 15°, 10° and 5°, respectively, to model the progressive readjustment of the molecules as expected in the experimental samples, with increasing the thickness. The results are summarized in Figure 5e: the bottom 4 molecules rotate and flatten, similarly to the central layer of the 20° model discussed above (panel 5b); the remaining units stay in the θ=20° configuration, as in the thick film. The resulting system has $\mu_T$=0.0 $\mu_B$ and $\mu_A$ =10.2 $\mu_B$/cell where the rotated molecules have zero magnetic moment, while the rest maintains the AF interactions. Therefore, in the general case, depending on the growth conditions, the films may show magnetic interactions different from the AF, due to the presence of uncompensated spin layers, whose number (i.e., the final magnetization) depends on the specific different stacking (i.e., intermolecular π−π coupling). The molecules in the samples grown at 300 K have sufficient kinetic energy to mimic the crystalline phase, while in samples grown at a lower temperature the molecules assemble adopting a different arrangement that gives rise to a XMCD signal, because of a spin unbalance in the molecular stack. Control of the growth parameters influence thin film growth and properties[41]: the substrate temperature during growth could be used as a tool to tune the magnetic character of purely organic thin films.

**Conclusions**

We measured the magnetic dichroism in thin films of purely organic radicals and found that the XMCD signature in the films is different than in the bulk depending on the film preparation. We explain this difference in terms of the different arrangement of the molecules in the two cases.

Even though this work presents two fundamental results, i.e., the magnetic character of purely organic radical thin films can be investigated using XMCD, and it is influenced by the preparation conditions, it is only a first step in a largely unknown field. Many more experiments are needed to catch the intimate nature of magnetic ordering in radical thin films: measuring the magnetization curves and the hysteresis loop versus the applied magnetic field to define a magnetic character, exploring a large set of preparation parameters and different radicals, including high-spin radical systems are the next necessary steps. New experimental approaches to the XMCD measurements suitable for purely carbon-based materials are also necessary: we suggest that, from the experimental point of view, in case of XAS at the absorption edges of light elements, measuring the clean substrate for minimizing the normalisation issues might be beneficial. In fact, in NEXAFS spectroscopy the best way to normalize the curves at the absorption edge of thin films of light elements such as carbon is using the clean substrate signal rather than the taking into account the total absorption of the beamline with a diode or a grid[42]. On the theoretical side, a model to interpret the XMCD results in this class of materials from a general point of view is missing. The present work will certainly stimulate more discussion in the field and provides exciting new avenues to a new generation of experiments, opening fascinating possibilities for magnetism, revealing the importance of synthesizing, chemically tuning and manipulating stable radicals, towards achieving collective magnetic behaviour in purely organic thin films, also at higher temperatures.

**Methods**

**Experimental section**. The molecules were synthetized as described in Ref.[8] Thin film were grown in UHV by using organic molecular beam deposition (OMBD). Native $SiO_2$ grown on single-side polished n-Si(111) wafer was used as the substrate for all thin films. The wafers were cleaned in an ultrasonic bath by immersion in ethanol and acetone for one hour each and annealed in ultra-high vacuum at around 500 K for several hours. The thin films were grown keeping the substrate temperature either at 290 or 300 K during deposition. When available, the cleanness of the substrates was checked by XPS. The Knudsen cell was accurately calibrated, and the same cell was used during all experiments. The cell temperature during evaporation was 418 K, far below the onset of degradation of the radical as determined with thermal gravimetric analysis.[8] The evaporation rate was 0.2 nm/min. Under these preparation conditions the films follow the Stranski-Krastanov growth mode (layer+islands). Previously to the experiments discussed int this work, the films were characterised by using electron paramagnetic resonance to prove that the chosen preparation conditions allow depositing films of intact radicals.[8,19] All measuring stations were equipped with preparations chambers that allowed installing the same calibrated Knudsen cells, therefore always using the same evaporation protocol. XPS measurements were performed in an UHV system consisting of a substrate preparation chamber, an OMBD-dedicated chamber, and an analysis chamber (base pressure 4 x $10^{-10}$ mbar) equipped with a SPECS Phoibos 150 hemispherical electron analyser and a monochromatic Al Kα source (SPECS Focus 500). Further information on substrate preparation, film growth, stoichiometry measurements and fit calculations are given in Ref. [8].

UPS and NEXAFS measurements were performed at the third-generation synchrotron radiation source Bessy II (Berlin, Germany) at the LowDose PES end-station, installed at the PM4 beamline ($E/\Delta E$=6000 at 400 eV) that included substrate preparation and film deposition facilities like those described above for the XPS station. The UPS measurements were carried out in single bunch mode with a SCIENTA ArTOF electron energy analyzer, the NEXAFS measurements in multibunch hybrid mode (ring current in top up mode=300 mA, $c_{ff}$=3, 100 μm exit slit). The NEXAFS spectra were measured using linearly polarised light, in normal and in grazing incidence (30° to the surface) and normalized by taking the substrate signal and the ring current into account[43].

The XMCD experimental setup[24] was installed at the XUV Beamlime P04 at the PETRA III synchrotron (Hamburg, Germany). At the Beamlime P04 each absorption edge could be scanned in a very short time, from few minutes to several seconds, and the beam exposure was minimized with a shutter. A magnetic field (± 7 T) was applied in the horizontal direction, collinear to the incoming photon beam. The measurements were carried out in normal incidence. Each element-resolved absorption curve was obtained averaging up to 54 scans, each measured on a fresh point on the film surface. All absorption spectra were recorded in total electron yield. To determine XMCD, the adsorption curves were divided by the incoming photon intensity current, $I_0$, measured with a photodiode, for the two polarisations. An offset was removed to align the c+ and the c- curves, making sure that the pre-edge regions are equal. We subtracted the same linear background to the c+ and the c- curves and we normalised to have the same equal edge jump. The c- and c+ curves obtained in this way were subtracted to calculate XMCD.[29,30]

We have taken all precautions necessary to avoid radiation damage (e.g., short time beam exposure, shutter protection, defocusing of the beam, low dose radiation). All samples were carefully monitored for radiation damage during beam exposure. All measurements were performed on freshly prepared films.

**Calculations**. Calculations were performed with the Quantum Espresso Package[44,45] which implements a planewave formulation of the Density Functional Theory (DFT). Exchange-correlation was treated in the Perdew–Burke–Ernzerhof (PBE) Generalized Gradient Approximation (GGA)[46], and the spin degrees of freedom were described within the local spin-density approximation (LSDA). Tests on the effect of the exchange and correlation functional and the comparison with hybrids are reported in SI. Ionic potentials were described by using ab initio ultrasoft pseudopotentials of Vanderbilt's type[47]. Single-particle electronic wave functions (charge) were expanded in a plane-wave basis set up to an energy cutoff of 28 Ry (280 Ry). Van der Waals corrections (Grimme formulation[48]) to dispersive forces were included to improve the description of the intra- and inter-molecular interactions. A uniform (4x8x2) k-point mesh was used for sampling the 3D Brillouin zone of the bulk; a (4x8) k-point mesh was used for the 2D Brillouin zone of the films.

The molecular crystal was simulated by using periodic supercell. The initial structure was extracted by X-ray experimental data.

The Blatter-pyr 3D crystal was simulated in a monoclinic unitary cell of dimensions a= 13.65 Å, b=5.11 Å, c=28.24 Å, a=90.00°, b=99.47°, g=90.00°, which includes four molecules as shown in Figure 4 and Supporting Information (Figure S4). The films were modelled through periodically repeated monoclinic slabs, each including 12 Blatter-pyr radicals and a vacuum layer (~1.5 nm) in the direction perpendicular to the surface plane, to avoid spurious interactions among the replicas. All structures were fully relaxed until forces on all atoms became lower than 0.03 eV/Å$^{-1}$.

NEXAFS spectra at the nitrogen K-edge were simulated by using the first-principles scheme based on the continued-fraction approach and ultrasoft pseudopotentials as implemented in the Quantum ESPRESSO package[49]. The reference 1s core level states were obtained by replacing the pseudopotential of a selected N atom with another one, which simulates the presence of a screened core hole. Since the molecule includes inequivalent nitrogen atoms, we repeated the NEXAFS calculations for each N atom of the radical and we averaged the resulting spectra. X-ray absorption spectra for each atom were also repeated changing the direction k$_i$ of incident electric field along the three cartesian axes. The results are summarized in the SI.

**Acknowledgements**

This paper is in memory of Wilfried Wurth. He was a brilliant scientist open to new ideas and views. We are deeply saddened that he will not see the result of our work because it was made possible also thanks to Wilfried's unprejudiced scientific approach. He would have appeared as a co-author.

The authors would like to thank Helmholtz-Zentrum Berlin (HZB) for providing beamtime at BESSY II (Berlin, Germany), and DESY (Hamburg, Germany), a member of the Helmholtz Association HGF, for the provision of experimental facilities at PETRA III, Tang Zahng, Moritz Hoesch, Kai Bagschik, Hilmar Adler, Elke Nadler and Sergio Naselli for technical support,

Thomas Chassé for the access to the photoelectron spectroscopy lab at the University of Tübingen. We also thank Eberhard Goering for the XAS normalisation method. Financial support from HZB, DESY, and German Research Foundation (DFG) under the contract CA852/5-2 and CA852/11-1 is gratefully acknowledged. We thank the National Science Foundation (NSF), Chemistry Division for support of this research under Grants No. CHE-1665256 (A.R.) and CHE-1955349 (A.R.).

**Author contributions**

T.J., I.B., M.G., F.K., M.N., J.S., M.M. and M.B.C. took part in the beamtimes at PETRA III. M.G., F.C., E.G., R.O. and M.B.C. took part in the beamtimes at BESSY II. N.N.G. and A.R. designed and synthetized the radical. A.C. performed the calculations. M.B.C. conceived and supervised the project; interpreted the data and wrote the manuscript together with A.C. All authors contributed to the discussion and commented on the manuscript.

**Competing interests** The authors declare no competing interests.

**Additional information**

**Supplementary information** is available for this paper at

**Reprints and permissions information** is available at http://www.nature.com/ reprints.

**Data availability** The datasets generated during and analysed during the current study are available from the corresponding author on reasonable request.

**Correspondence and requests for materials** should be addressed to benedetta.casu@uni-tuebingen.de.

a)

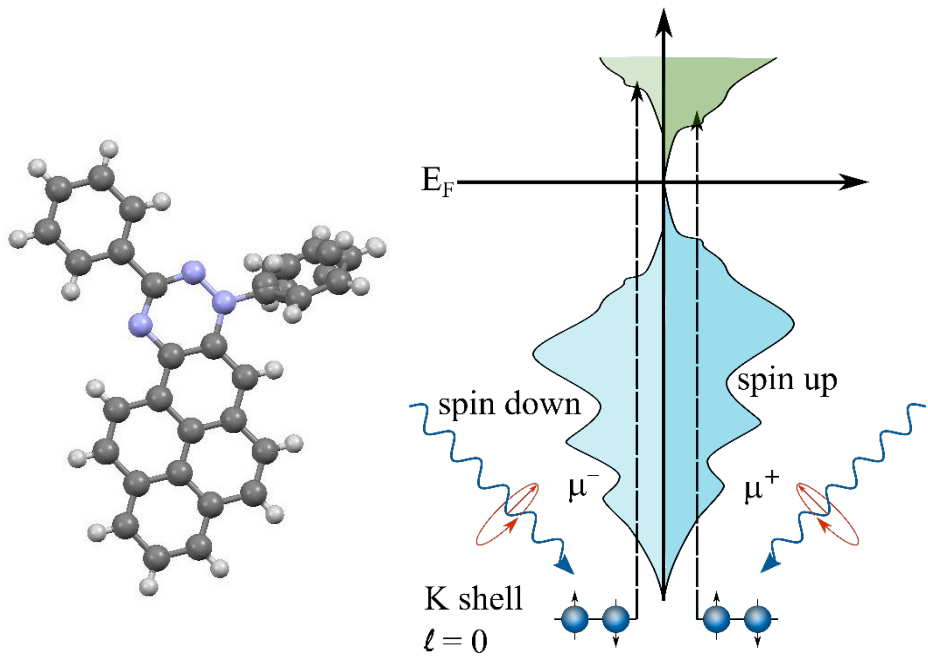

b)

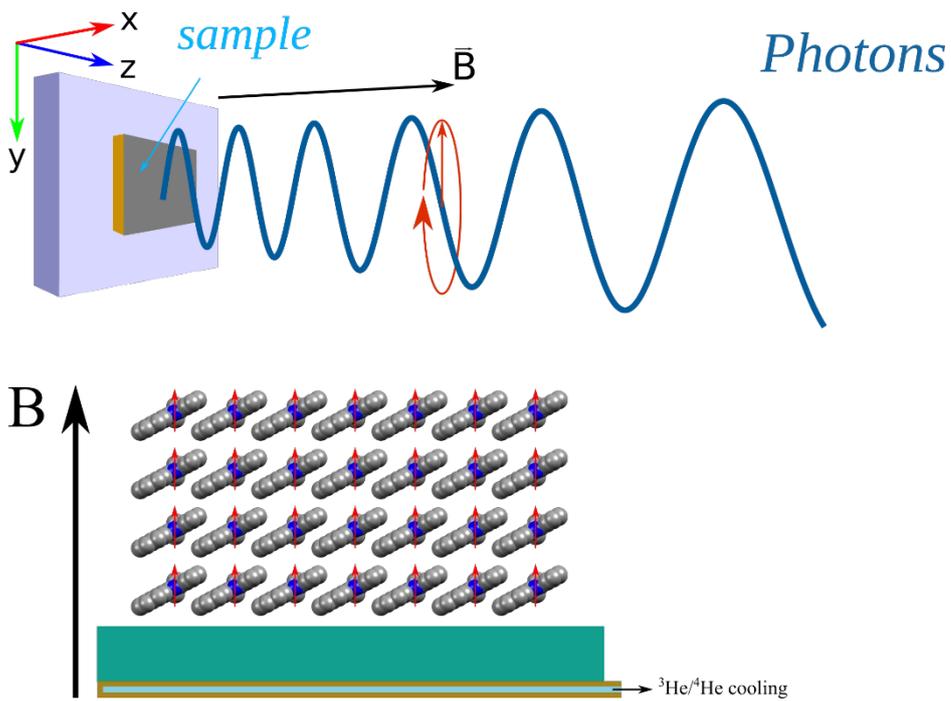

**Figure 1 | Molecular structure of the pyrene Blatter derivative and experiment. a**. (left panel) Molecular structure ($C_{29}H_{18}N_3^\bullet$), carbon atoms in dark grey, nitrogen atoms in blue, and hydrogen atoms in white. (right panel) XMCD principles. **b**. Sketch of the experiment.

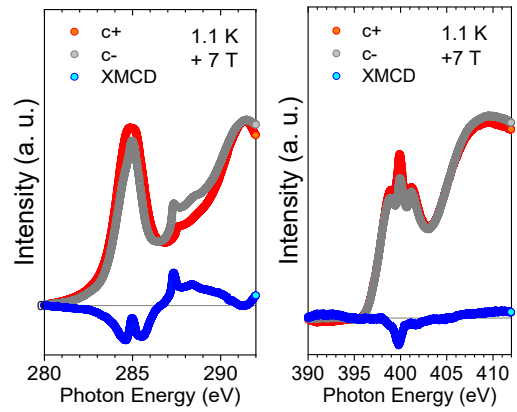

**Figure 2 | X-ray absorption measurements.** Circularly polarized (left) C K-edge and (right) N- K-edge XAS and XMCD as indicated recorded in normal incidence at B=7 T, at 1.1 K (substrate temperature during preparation at 290 K, nominal thickness 9 nm).

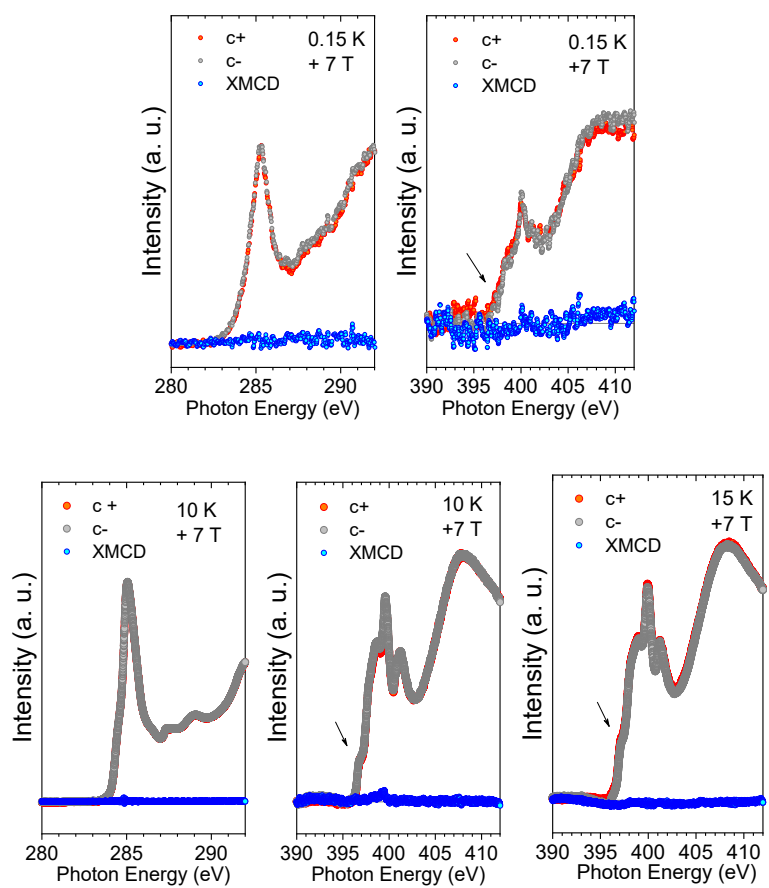

**Figure 3 | X-ray absorption measurements.** Circularly polarized C K-edge and N- K-edge XAS and XMCD recorded in normal incidence at B=7 T, and temperature as indicated (substrate temperature during preparation at 300 K, nominal thickness 4 (upper panel) and 9 nm (lower panel)). The spectra at 0.15 K were measured with lower resolution. The features are therefore less resolved than at 10 K but still present (see also Supplementary information, Figure S 3)

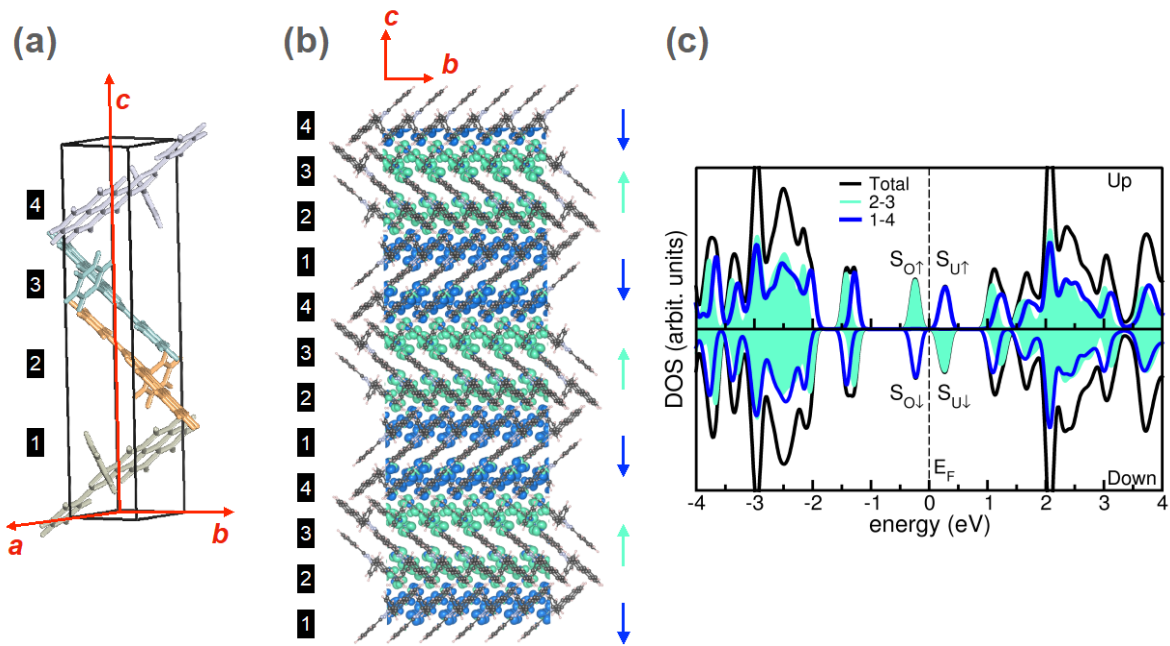

**Figure 4 | Calculation results for the Blatter-pyr antiferromagnetic film. a.** Primitive cell of monoclinic Blatter-pyr crystal. Labels 1-4 indicate the four molecules in the cell. **b.** Side view of the 9 nm-thick film. Coloured isosurfaces represent the charge spin-density plots for the antiferromagnetic phase. Green (blue) colour indicates the majority (minority) spin-up (spin-down) orientation of single radicals. Atomic positions and charge spin-density have been replicated for clarity. **c.** Total and molecule projected spin-polarized density of states. Total (black line) and molecule projected (shaded green areas for molecules 2-3 and thick blue line for molecules 1-4) spin-polarized density of states (DOS) of the Blatter-pyr film. Vertical dashed line marks the Fermi level of the system assumed as energy reference for the plot. $S_O$ ($S_U$) indicate the single occupied (unoccupied) molecular orbitals for the spin up (↑) and spin down (↓) channels.

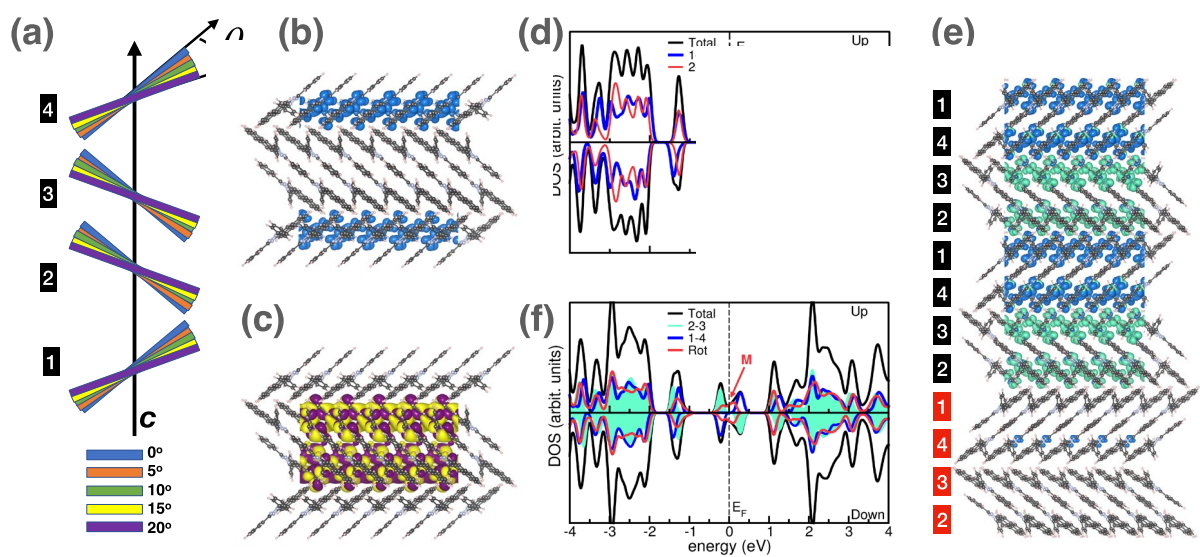

**Figure 5 | Calculation results for different molecular orientations in the Blatter-pyr films.** **a.** Graphical model for a 3 nm-think film (4 molecules), rotated by angle θ with respect to the crystalline *c*-axis. Colours identify different rotation angles, θ=0° corresponds to the crystalline configuration (no rotation). **b.** Charge spin density plot of rotated model with θ =20°. **c.** Isosurface plot of the mixed M state and **d.** spin polarized DOS corresponding to the model in panel c. **e.** Side view and charge spin density plot for partially 9-nm thick film with a different molecular arrangement at the bottom. Red labels correspond to rotated molecules (20°, 15°, 10°, 5° for molecule 2, 3, 4, 1 respectively); black labels correspond to un-rotated stacked molecules. (f) Spin-polarized DOS corresponding to the structure in panel e). Red lines in panels c and f correspond to the projected DOS from spin-unpolarized molecules.